\documentclass[conference,a4paper]{IEEEtran}
\usepackage{amssymb}
\usepackage{color}
\usepackage{graphicx}
\usepackage{amsmath}
\usepackage{cite}
\usepackage{pgfplots}
\usepackage{comment}
\usepackage{multirow}
\usepackage{mathbbol}
\usepackage{caption}
\usepackage{booktabs}
\usepackage{caption}
\usepackage{subfig}
\usepackage{flushend}

\def\C#1{\mathcal{#1}}

\pgfplotsset{compat=newest}

\pgfplotsset{mystyle/.style={%
        width=6cm,
        xmin=0,xmax=0.5,
        xtick={0,10,...,50}}}

\newcommand{\E}[1]{\mathop{\mathbb E}\nolimits\left[#1\right]} 

\newcommand{\ie}{i.e.,\,}

\newcommand{\figuresname}[1]{Figs.~}

\newcommand{\NVB}{\textcolor{blue}{$\Rightarrow$}\begin{rm} \color{blue}}   
\newcommand{\NVE}{\end{rm}\textcolor{blue}{$\Leftarrow$}}

\newif\ifcutshort
\cutshortfalse
\newcommand{\quotes}[1]{``#1''}

\newcommand{\e}[1]{%
	\ifmmode\refstepcounter{equation}%
	  \eqno\mbox{\rm(\theequation)}\label{e:#1}%
	\else(\ref{e:#1})\fi}

\catcode`@=11
\newdimen\jot \jot=3pt
\def\openup{\afterassignment\@penup\dimen@=}
\def\@penup{\advance\lineskip\dimen@
  \advance\baselineskip\dimen@
  \advance\lineskiplimit\dimen@}
\def\eqalign#1{\null\,\vcenter{\openup\jot\m@th
  \ialign{\strut\hfil$\displaystyle{##}$&$\displaystyle{{}##}$\hfil
      \crcr#1\crcr}}\,}

\graphicspath{{./figure/}}

 \usepackage[acronym]{glossaries}
 \newacronym{tdma}{TDMA}{Time Division Multiple Access}
\newacronym{ack}{ACK}{Acknowledgment}
\newacronym{dc}{DC}{Duty Cycle}
\newacronym{dr}{DR}{Data Rate}
\newacronym{dl}{DL}{downlink}
\newacronym{cca}{CCA}{Clear Channel Assessment}
\newacronym{dcw}{DCW-MAC}{Duty-Cycled Medium Access Scheme for Low-Power WUR}
\newacronym{awd}{AWD-MAC}{Asynchronous Wake-up on Demand MAC}
\newacronym{lecim}{LECIM}{Low-energy Critical Infrastructure Monitoring}
\newacronym{scm}{SCM-WUR}{Sub-Carrier Modulation Wake-up Radio Protocol}
\newacronym{opwum}{OPWUM}{Opportunistic Wake-Up MAC Protocol}
\newacronym{ed}{ED}{End Device}
\newacronym{en}{EN}{End Node}
\newacronym{rtt}{RTT}{Round Trip Time}
\newacronym{gw}{GW}{Gateway}
\newacronym{iot}{IoT}{Internet of Things}
\newacronym{ism}{ISM}{Industrial, Scientific, and Medical}
\newacronym[plural=LPWANs,firstplural=Low Power Wide Area Networks (LPWANs)]{lpwan}{LPWAN}{Low Power Wide Area Network}
\newacronym{mcs}{MCS}{Modulation and Coding Scheme}
\newacronym{mac}{MAC}{Medium Access Control}
\newacronym{pcr}{PCR}{Primary Communication Radio}
\newacronym{phy}{PHY}{Physical}
\newacronym{pmf}{PMF}{Probability Mass Function}
\newacronym{aoi}{AoI}{Age of Information}
\newacronym{qos}{QoS}{Quality of Service}
\newacronym{wsn}{WSN}{Wireless Sensor Network}
\newacronym{ul}{UL}{uplink}
\newacronym{wur}{WUR}{Wake-Up Radio}

\def \fwidth{0.95\columnwidth}
\def \fheight {0.6\columnwidth}

\definecolor{color1}{HTML}{FFB14E}
\definecolor{color2}{HTML}{FA8775}
\definecolor{color3}{HTML}{EA5F94}
\definecolor{color4}{HTML}{CD34B5}
\definecolor{color5}{HTML}{9D02D7}
\definecolor{color6}{HTML}{0000FF}

\begin{document}

\title{Low-Latency Massive Access with Multicast Wake Up Radio}

\author{\IEEEauthorblockN{Andrea Zanella, Anay Ajit Deshpande, Federico Chiariotti}
\IEEEauthorblockA{Department of Information Engineering, University of Padova\\
Via G. Gradenigo 6/B, Padova, Italy\\
Email: \texttt{\{zanella,deshpande,chiariot\}@dei.unipd.it}}}

\maketitle


\begin{abstract}
The use of \gls{wur} in \gls{iot} networks can significantly improve their energy efficiency: battery-powered sensors can remain in a low-power (sleep) mode while listening for wake-up messages using their \gls{wur} and reactivate only when polled, saving energy. However, polling-based \gls{tdma} may significantly increase data transmission delay if packets are generated sporadically, as nodes with no information still need to be polled. In this paper, we examine the effect of multicast polling for \gls{wur}-enable wireless nodes. The idea is to assign nodes to multicast groups so that all nodes in the same group can be solicited by a multicast polling message. This may cause collisions, which can be solved by requesting retransmissions from the involved nodes. We analyze the performance of different multicast polling and retransmission strategies, showing that the optimal approach can significantly reduce the delay over \gls{tdma} and ALOHA in low-traffic scenarios while keeping good energy efficiency.
\end{abstract}


\section{Introduction}
\glsresetall

The explosion of the \gls{iot} has led to new developments in remote monitoring applications~\cite{wang2021evolution}, which use distributed sensors to track, e.g., environmental conditions in remote and/or wide areas \cite{zanella2023iot}, as well as manufacturing plants and cities. Since the inception of the \gls{iot}, however, energy has been a major issue for system design~\cite{georgiou2017iot}: battery-powered nodes face significant constraints in terms of computational and communication capabilities, and often resort to uncoordinated random access schemes like ALOHA to avoid signaling overhead.

However, the limits of random access schemes are well-known \cite{8030484}: unless the traffic is extremely light, these schemes suffer from packet collisions and congestion~\cite{yu2020stabilizing}, and may not allow the \gls{gw} to request new data from a specific sensor. In contrast, polling schemes avoid collisions, but can result in significant energy consumption for nodes that must constantly listen for request messages, without constraining scheduling to a fixed duty cycle. \cite{levy1990polling}. 

One possible solution to this problem is provided by \gls{wur} technology, standardized by the IEEE as part of the 802.11 group~\cite{deng2020ieee}: an extremely low-power radio, only capable of receiving simple signals and performing basic calculations, which can be kept continuously active in listening mode, while the sensor's main processor and \gls{pcr} are only turned on when needed. Typically, the \gls{wur} is used to reduce the downlink response time of a node whose \gls{pcr} is in sleep mode to save energy\cite{tang2017energy}, potentially reducing both latency and energy consumption.

However, polling one sensor at a time can still cause significant delays between the moment a measurement is generated and when it is received by the \gls{gw}~\cite{rostami2019wake}. As the latter does not know when sensors have new measurements, it can employ smart strategies based on data importance or \gls{aoi}~\cite{chiariotti2022scheduling}, but it will still waste a significant amount of time polling sensors with no data to transmit if the average traffic is low. This problem is well-known, and shared with traditional \gls{tdma}, as even advanced scheduling schemes cannot know in advance whether a specific sensor has a new packet~\cite{shiraishi2022query}.

In this work, we propose a solution that, by exploiting the standard \gls{wur} features, can combine the desirable properties of polling and random access schemes: instead of polling sensors one at a time, the \gls{gw} sends multicast polling messages to groups of sensors, resolving possible collisions only if and when they occur. The protocol includes two components: first, the nodes-grouping strategy, which is optimized to limit the collision probability while still exploiting the channel as effectively as possible, and second, the collision resolution mechanism. The latter is crucial to avoid the instability of ALOHA: if resolving a collision takes too long, the other sensors may accumulate a backlog, increasing the probability of further collisions and forcing the \gls{gw} to poll smaller groups or risk making the system unstable.

The stability of the proposed protocol and its performance in terms of average delay and energy efficiency are analyzed via Monte Carlo simulation, which shows that multicast polling can outperform both ALOHA and \gls{tdma} in light traffic conditions, at the cost of slightly lower energy efficiency. In particular, the proposed method can reduce the delay by 90\% in a network of 100 nodes, and by 99\% in a massive access scenario with 1000 sensors, with respect to both ALOHA and unicast \gls{wur} polling. At the same time, the energy consumption for the sensors is higher, but never more than twice that of \gls{tdma}, which is the most energy-conservative scheme among the ones we analyzed.

The rest of this paper is organized as follows: Sec.~\ref{sec:related} presents a review of the literature on \gls{wur}-based \gls{mac} schemes, highlighting their advantages and limitations. We then present the basic system model and our multicast polling protocol in Sec.~\ref{sec:system}, and provide some analytical results on its performance in Sec.~\ref{sec:analysis}. Full results in a more realistic setting are provided and discussed in Sec.~\ref{sec:results}. Sec.~\ref{sec:conc} concludes the paper and presents some possible avenues of future work on the subject.

\section{Related Work}\label{sec:related}
\gls{mac} protocols for \glspl{wsn} have been investigated quite thoroughly in the literature, and hundreds of schemes and variations have been proposed over the decades. However, classic protocols such as ALOHA are still used in a dominant share of \glspl{wsn} due to their simplicity and low overhead. 

Another common design philosophy involves the use of polling schemes where the \gls{gw} grants dedicated channel access to one node at a time.  This reduces complexity and avoids interference, but requires the nodes to constantly listen for polling messages, thus increasing energy consumption~\cite{feeney2001investigating}. \gls{wur} can alleviate this problem, being explicitly designed to consume very little power in the listening state. For this reason, several \gls{wur}-based \gls{mac} protocols have been recently proposed\cite{mahlknecht2009wur}. 

\subsection{The IEEE 802.11ba Standard}
The IEEE 802.11ba is a proposed extension to legacy IEEE 802.11 standard~\cite{bankov2019ieee} with the purpose of enhancing power efficiency in battery-powered \gls{iot} nodes. The IEEE802.11ba standard is primarily focused on the inclusion of a low-power \gls{wur}, which is used to listen for polling signals and activate the \gls{pcr}. The standard defines the physical and \gls{mac} parameters for communication with the \gls{wur}, as well as the wake-up procedure for the \gls{pcr} when a \gls{wur} signal is received from the \gls{gw}. It also includes the power management scheme to be implemented, the \gls{dc} specification for the \gls{wur}, and a synchronization scheme using \gls{wur} beacons. Finally, the crucial component for the mechanism proposed in this paper is the channelization of wake-up frames to be sent to \gls{wur}. The standard explains the usage of group cast wake-up messages to be sent to multiple nodes, waking up their \glspl{pcr} simultaneously. However, it does not mention the size or the periodicity of the messages to be sent out. This provides the basis for the design of our mechanism.

\begin{figure}[!t]
    \centering
    \includegraphics[width=0.95\columnwidth]{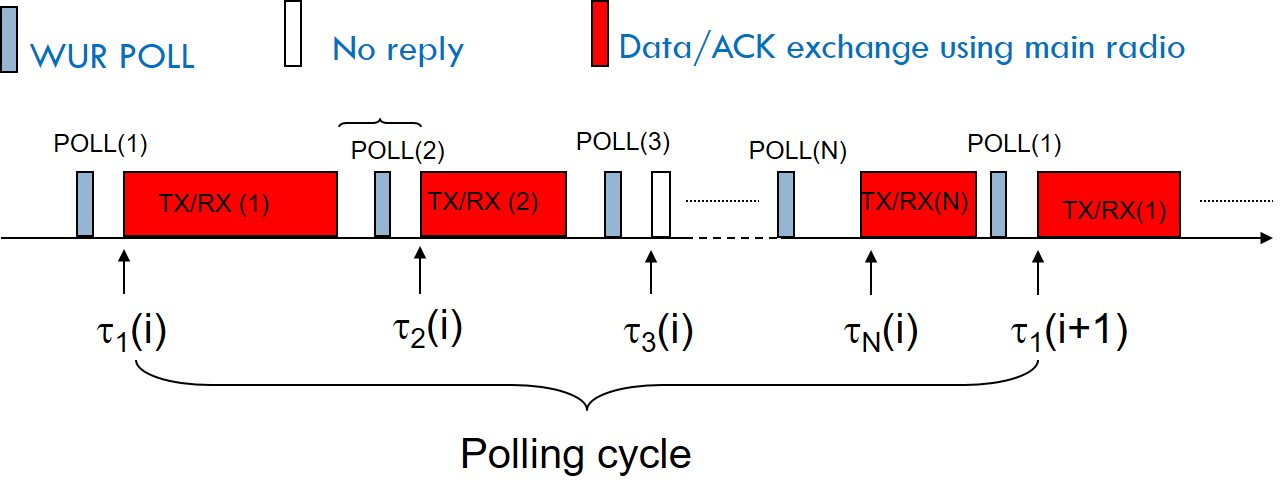}
    \caption{WUR-based simple polling scheme for all the nodes in the network}
    \label{fig:poll}
\end{figure}

\subsection{WUR MAC Protocols}

\gls{wur}-based polling protocols have been a topic of research even before the advent of IEEE 802.11ba. Polling protocols that have been proposed so far use \gls{wur} as a listening device for \gls{cca} or act as a trigger to switch on the \gls{pcr}. In \gls{dcw}~\cite{mazloum2011dcw}, the \gls{wur} receiver is periodically turned on by the node to listen to a wake-up message in order to activate the main radio for data transmission. Due to the reduced listening time of \gls{wur}, the scheme consumes extremely little power but may experience a significant delay in case the wake-up message transmission has to be repeated until it coincides with a \gls{wur} listening period. In \gls{awd}~\cite{le2015asynchronous}, the node sends out a broadcast beacon using its \gls{wur} to advertise that it is ready to receive. Upon receiving the beacon, the transmitter turns on its \gls{pcr} and transmits an ACK message. Once the receiver gets the ACK message on its \gls{wur}, it turns on its \gls{pcr} for data transmission. In~\cite{ullah2012energy}, a \gls{mac} protocol is introduced for \gls{lecim}. The protocol works on slotted ALOHA with the use of \gls{wur}. For uplink communication, the \gls{en} activates its \gls{wur} and listens to the channel for beacon frames to synchronize with the \gls{gw}. Once it receives the beacon packet, it synchronizes with the superframe of slotted ALOHA and transmits over a randomly chosen slot. If the transmission is successful, the node goes to sleep mode, otherwise, it transmits over another random slot. \gls{opwum} \cite{ait2016opwum} is a multi-hop protocol that uses the \gls{wur} to broadcast a relay request to all the neighbors of a node with packets to transmit. The nodes reply with \glspl{cca} acknowledgment, still using the \gls{wur}. Based on the shortest backoff timer from the clear channel messages, the original transmitter then chooses the next relay node among its neighbors. In \gls{scm}\cite{oller2015has}, the transmitter broadcasts a wake-up message to all nodes in  its coverage range containing a unique identifier of the intended receiver, which is the only one that activates its \gls{pcr} and listens to the channel for successive data transmissions. After the data exchange, both transmitter and receiver switch off their \glspl{pcr}, but keep listening to the channel using their \glspl{wur}.

The protocols proposed so far, especially for the use of \gls{wur} in uplink communication, employ either additional mechanisms such as \gls{cca} to determine whether the channel is free to transmit or employ ALOHA-based schemes, with a higher chance of collisions. Both these families of schemes have longer delays and consume more energy than \gls{tdma} in heavy traffic conditions because of channel sensing or retransmissions. However, \gls{tdma} may waste a lot of time in polling nodes with no new data when the traffic over the network is light. In order to overcome this problem, we propose a centralized multicast polling scheme, which not only takes into account the nodes that are most likely to have data to transmit, but also the time duration between the polls to each node. This system helps in reducing the delay considerably while maintaining a limited collision probability.

\section{System Model and Protocol Definition}\label{sec:system}
In this section, we first investigate a basic polling mechanism using \gls{wur} and then propose an extension to implement the group polling introduced in the previous section. 
Let us consider a \gls{wsn} with a single \gls{gw} serving $N$ sensors. Each sensor $n$ generates new updates according to an independent Poisson process with rate $\lambda_n$. We will say that a node is \quotes{activated} when it has a packet to transmit. We assume a pure collision channel model, i.e., if any part of a transmission collides with another, both packets are lost. For simplicity's sake, we also assume the wake-up time for the \gls{pcr} to be constant, and that wake-up messages are never lost. In addition, we assume the \gls{pcr} transmits a fixed data rate, and packets have constant size. Although these assumptions can be trivially relaxed with minimal impact on the system performance, this would make the model more cumbersome to define and analyze. 

In the traditional round-robin polling scheme, the \gls{gw} is tasked with polling every single sensor in the network in a predetermined and fixed order. If the solicited sensor has any new data, it turns on its \gls{pcr} and transmits a packet otherwise it remains silent. If the \gls{gw} receives the reply packet, it will immediately send an ACK and then proceed with the polling cycle. Otherwise, if the polled node does not send any packet, the \gls{gw} will resume the polling just after a pre-defined time interval, which depends on the maximum \gls{rtt} of the \gls{wsn}, that is much shorter than a packet transmission time. \figurename~\ref{fig:poll} shows a simple schematic of the protocol. More advanced schemes, which reduce the guard interval depending on the distance between sensor and \gls{gw} are also possible, but the basic structure of the protocol will remain the same. This polling scheme follows the IEEE 802.11ba standard, with in-band \gls{wur} wake-up signals.

\subsection{Polling slot duration}
We define as polling slot the time interval between two consecutive polling packets of the \gls{gw}. The duration of a polling slot then depends on the amount of data that the polled node needs to return to the \gls{gw}. Let $\tau_0$ denote the maximum signal propagation time from the \gls{gw} to the furthest node and return. In case of no data to be transmitted (in which case, we say the slot is \textit{idle}) the duration is $\tau_0+t_{\text{wu}}$, where $t_{\text{wu}}$ is the duration of a wake-up signal~\cite{hwang2018wake}. Instead, if the sensor has generated $k$ new packets since the last update, the polling slot duration becomes $kt_p+t_{\text{wu}}+\tau_0 $ where $t_p$ is the transmission time for a single packet. Therefore, for any $k\in\mathbb{N}$, the \gls{pmf} of the slot duration $t_n$, when sensor $n$ is polled, is  given by
\begin{equation}
   p_{t_n}(kt_p+t_{\text{wu}}+\tau_0)=\frac{(\lambda_n\tau_n)^{k}e^{-\lambda_n\tau_n}}{k!}, 
\end{equation}
and zero otherwise, where $\tau_n$ is the duration of the polling cycle for the $n$ sensor, i.e., the total time between two polling slots of that node.  

\subsection{Multicast polling scheme}
The energy efficiency of the \gls{wur} unicast polling scheme is significantly higher than for standard \gls{tdma} or legacy polling through the \gls{pcr}, as the nodes do not need to be synchronized or stay awake all the time. However, the delay of such a scheme is high, as all nodes need to be polled and new data is often queued for a relatively long time, even if the traffic is very low. This problem is an open issue in the 802.11ba standard.
The legacy alternative, i.e., a simple ALOHA scheme, is energy efficient, as sensors only wake up when they have data, but the reliable transmission requirement makes it problematic even at very low traffic loads: any collision causes retransmission after a random back-off time, increasing the delay and energy consumption, and the well-known theoretical load limit of ALOHA is below 20\%.

In order to overcome this trade-off, we propose a multicast polling scheme: instead of polling a single node at a time, the \gls{gw} polls multiple sensors. If more than one polled node has data to transmit, the packets collide, and the \gls{gw} uses a collision resolution scheme to poll the nodes separately. Naturally, the larger the group, and the higher the traffic load within the group, the higher the probability of collision, making the selection of the group size and collision resolution scheme critical for performance. Collisions affect both delay and energy consumption, as resolving a collision takes time, and each transmission, whether successful or collided, requires the same amount of energy.

If we consider the selection of a group $\C{G}$ of $G=|\C{G}|$ sensors, the transmission is successful if and only if exactly one sensor has new data to transmit. If we denote $\tau_n$ as the time since the end of the last polling slot for sensor $n$, we can easily compute the probability of exactly one node transmitting, i.e., the success probability for group $\C{G}$ as
\begin{equation}
    p_s(\C{G})=\sum_{i\in\C{G}}\left(1-e^{-\lambda_i\tau_i}\right)\exp\left(-\sum_{j\in\C{G}\setminus\{i\}}\lambda_j\tau_j\right).
\end{equation}
where the first term in the summation gives the probability that node $i$ has collected at least one packet since the last transmission, while the rightmost term is the joint probability that all the other nodes in the same group have no backlog. The probability that the slot will be idle is given by
\begin{equation}
    p_{\text{idle}}(\C{G})=\exp\left(-\sum_{i\in\C{G}}\lambda_i\tau_i\right);
\end{equation}
and the collision probability is then
\begin{equation}
p_c(\C{G})=1-p_{\text{idle}}(\C{G})-p_s(\C{G}).\label{eq:coll}
\end{equation}

\subsection{Nodes grouping}
Naturally, if $G=1$, $p_c(\C{G})=0$. However, in the general case, the success and collision probabilities depend not only on the group size but on its composition as well: finding the group with the highest success probability is a combinatorial problem similar to the well-known knapsack problem.

\begin{figure}[!t]
    \centering
    \includegraphics[width=0.95\columnwidth]{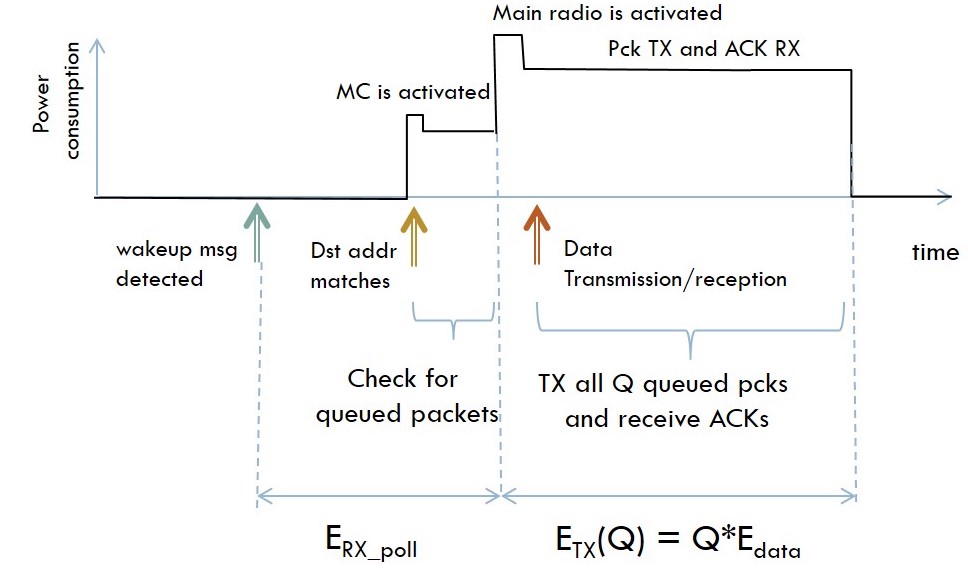}
    \caption{Energy Consumption based on the each step of activation and transmission using WUR and PCR}
    \label{fig:comp2}
\end{figure}

We assume that the \gls{gw} knows the packet generation rate of all the sensors. Since it also knows $\tau_n$ for each sensor in the network, it can compute the collision probability for any set $\C{G}$. We then propose a simple constructive heuristic for node selection, which considers a maximum target collision probability $p_{\text{thr}}$: first, we sort the nodes in decreasing order of their probability of having packets to be transmitted. The node with the highest probability is the first to be inserted into the polling group. Then, we progressively add other nodes with the \textit{lowest} probability of having packets to be transmitted until the collision probability computed using~\eqref{eq:coll} reaches the prefixed target (or all nodes are in the group). The rationale is to group together one node with high transmission chances and a bunch of nodes with small transmission probabilities, so as to decrease the risk of idle polling, while also limiting the size of colliding nodes in case of multiple transmissions. Once the group has been polled, the procedure is repeated anew to form the next group. 

Higher values of the target threshold will lead to larger groups, with a lower probability of the slot remaining idle; however, an increased collision probability will also cause further delays. Although group-based polling is more robust than ALOHA, our results will show that increasing the load on the network will lead the algorithm to fall back on unicast polling, as the added delay may cause the system to become unstable.

\subsection{Collision resolution}
When collisions do happen, however, we have two choices to ensure full reliability. The first and simplest is to perform a linear search, polling all sensors in the group one by one. However, this requires a relatively long time, particularly if the group is large, and new packets may arrive in the meantime. If we denote the time required to complete the collision resolution process as $T_c(\C{G})$, we have
\begin{equation}
    T_c(\C{G})\geq G\left(\tau_0+t_{\text{wu}}\right)+2t_p,
\end{equation}
where the right-hand side of the inequality represents the minimum possible value obtained when only two packets collide, and no other packets arrive during the collision resolution process. This process is long, particularly when $G$ is large. Binary search can be more efficient. In case of collision, the group is split into two subgroups with similar aggregate activation probability, and each part is polled independently. The process is repeated recursively until all collisions are resolved. This requires some extra energy consumption, as multiple iterations might be needed to solve the collision, but as long as traffic is relatively sparse, the time required for the collision resolution is $O(\log_2(G))$ instead of $O(G)$, leading to significant benefits for larger groups.

\section{Performance Analysis}\label{sec:analysis}

We can now analyze the performance of unicast and multicast \gls{wur}. The performance of pure ALOHA in this scenario is not included in the paper, as it is a classical medium access scheme, and several in-depth analyses are available in the relevant literature.
\subsection{Unicast WUR}

Unicast \gls{wur}, which we also refer to as \gls{tdma}, has no collisions. Consequently, we can easily compute the duration of a polling cycle. Each wake-up signal requires $\tau_0+t_{\text{wu}}$, and each packet requires time $t_p$, so if $K$ packets are transmitted during a cycle, the total time is $T_u(K)=N\left(\tau_0+t_{\text{wu}}\right)+Kt_p$. However, the longer the duration of a cycle, the more packets are generated, so the average cycle duration is the solution of the following linear system:
\begin{equation}
    \begin{cases}
        &\mathbb{E}[K]=\sum_{n=1}^N\lambda_N \mathbb{E}[T_u];\\
        &\mathbb{E}[T_u]=N\left(\tau_0+t_{\text{wu}}\right)+\mathbb{E}[K]t_p.
    \end{cases}
    \label{eq:polling_cycle}
\end{equation}
The average duration of a polling cycle is then given by:
\begin{equation}
    \mathbb{E}[T_u]=\frac{N\left(\tau_0+t_{\text{wu}}\right)}{1-\sum_{n=1}^N\lambda_N t_p}.
\end{equation}
Naturally, the system does not work if the load is higher than 1, i.e., if $\sum_{n=1}^N\lambda_N \mathbb{E}[T_d]\geq 1$, as the polling cycle duration will become infinite. As packets may arrive at any time during the cycle, the average delay $D_u(n)$ for sensor $n$ is then simply given by:
\begin{equation}
 D_u(n)=\frac{\mathbb{E}[T^2_u]}{\mathbb{E}[T_u]}\frac{(1+\lambda_n t_p)}{2}.
 \label{eq:delay_eq}
\end{equation}
We can compute $\mathbb{E}[T^2_u]$ as
\begin{equation}
\begin{aligned}
    \mathbb{E}[T^2_u] &= \mathbb{E}[(N(t_{wu}+\tau_0)+Kt_p)^2],\\
    &= N^2(t_{wu}+\tau_0)^2 + 2N(t_{wu}+\tau_0)t_p \mathbb{E}[K] + t_p^2\mathbb{E}[K^2].
    \label{eq:polling_cycle_calc}
\end{aligned}
\end{equation}
As the number of packets $K$ follows a Poisson distribution, and defining symbol $\Lambda = \sum_n^{N-1}\lambda_n$, we can write $\mathbb{E}[K] = \Lambda\mathbb{E}[T_u]$ and $\mathbb{E}[K^2] = \Lambda\mathbb{E}[T_u] + \Lambda^2(\mathbb{E}[T_u])^2$. Substituting the values of $\mathbb{E}[K]$ and $\mathbb{E}[K^2]$ in \eqref{eq:polling_cycle_calc}, we get:
\begin{equation}
\begin{aligned}
    \mathbb{E}[T_u^2] =& N^2(t_{wu}+\tau_0)^2 + 2N(t_{wu}+\tau_0)\Lambda\mathbb{E}[T_u]\\
    &+t_p^2\Lambda\mathbb{E}[T_u]+t_p^2 \Lambda^2(\mathbb{E}[T_u])^2.
    \label{eq:tu_expectation}
\end{aligned}
\end{equation}
Substituting \eqref{eq:tu_expectation} in \eqref{eq:delay_eq}, we get:
\begin{equation}
\begin{aligned}
    D_u(n) =& \Bigg[N(t_{wu}+\tau_0)\Bigg(1-t_p\Lambda + 2Nt_p\Lambda + \\ &\frac{t_p^2\Lambda^2}{1-t_p\Lambda}\Bigg)+t_p^2\Lambda\Bigg]\frac{(1+t_p\lambda_n)}{2}.
\end{aligned}
\end{equation}
We can then consider the energy consumption of the protocol, which is the sum of several steps as shown in \figurename~\ref{fig:comp2}. We assume that the energy consumed by the \gls{wur} is negligible (being two or three orders of magnitudes lower than that of the \gls{pcr}). Each sensor then consumes energy $E_{\text{wu}}$ every time it receives a wake-up message, whether it is meant for it or not. If the address matches and the sensor has new data to transmit, the \gls{pcr} is turned on to transmit the messages and receive the corresponding ACKs.  Using cumulative ACKs can reduce energy consumption when transmitting multiple packets, but the protocol becomes slightly more involved. For simplicity of explanation, in the following we assume that each packet is individually ACKed. Therefore, denoting by $E_{\text{tx}}$ the energy required to transmit one packet and receive the corresponding ACK, a node with $k$ packets in its transmission buffer will spend a total energy of $kE_{\text{tx}}$. The average power consumption of a node $n$ is then given by:
\begin{equation}
  P_u=\frac{NE_{\text{wu}}}{\mathbb{E}[T_u]}+\lambda_nE_{\text{tx}}.
\end{equation}
The energy consumption $E_u$ for each successfully transmitted packet is hence equal to:
\begin{equation}
  E_u=\frac{P_u}{\lambda_n} = \frac{NE_{\text{wu}}}{\lambda_n\mathbb{E}[T_u]}+E_{\text{tx}}.
\end{equation}

\begin{figure*}[!t]
\centering
\subfloat[%
  Single Node \label{fig:cdf_single}%
]{
%
%

\begin{tikzpicture}

\begin{axis}[%
width=\fwidth,
height=\fheight,
xmin=0,
xmax=12,
xlabel style={font=\color{white!15!black}},
xlabel={Number of Rounds},
ymin=0,
ymax=1,
ylabel style={font=\color{white!15!black}},
ylabel={$P_{s}$},
xmajorgrids,
ymajorgrids,
legend style={legend cell align=left, at={(1.0,0.28)}, legend columns=1, align=left, draw=white!15!black}
]
\addplot [color=color5, line width=1.5pt, mark=square, mark options={solid}]
  table[row sep=crcr]{%
-inf 0\\
3	0\\
3	0.470897484040556\\
4	0.470897484040556\\
4	0.726248591813744\\
5	0.726248591813744\\
5	0.864814119414195\\
6	0.864814119414195\\
6	0.930529478032294\\
7	0.930529478032294\\
7	0.960195268494179\\
8	0.960195268494179\\
8	0.978971085242208\\
9	0.978971085242208\\
9	0.991363124295907\\
10	0.991363124295907\\
10	0.996620352985355\\
11	0.996620352985355\\
11	0.998873450995118\\
12	0.998873450995118\\
12	1\\
inf	1\\
};
\addlegendentry{Monte Carlo}
\addplot [color=color3, line width=1.5pt, mark=triangle, mark options={solid}]
  table[row sep=crcr]{%
1	0.0009331878021845\\
2	0.0554689157775412\\
3	0.261378310544736\\
4	0.522879109380922\\
5	0.726913343560756\\
6	0.853676939271055\\
7	0.924235495602606\\
8	0.961446275075031\\
9	0.980552615128738\\
10	0.990233341171281\\
11	0.995105886760501\\
12	0.997550242129724\\
};
\addlegendentry{Theoretical}
\end{axis}

\end{tikzpicture}%
}
\subfloat[%
  All Nodes \label{fig:cdf_all}%
]{
%
%
\begin{tikzpicture}

\begin{axis}[%
width=\fwidth,
height=\fheight,
xmin=0,
xmax=12,
xlabel style={font=\color{white!15!black}},
xlabel={Number of Rounds},
ymin=0,
ymax=1,
ylabel style={font=\color{white!15!black}},
ylabel={$P_{sa}$},
xmajorgrids,
ymajorgrids,
legend style={legend cell align=left, at={(1.0,0.51)}, legend columns=1, align=left, draw=white!15!black}
]
\addplot [color=color6, line width=1.5pt, mark=diamond, mark options={solid}]
  table[row sep=crcr]{%
-inf	0\\
3	0\\
3	0.487341772151899\\
4	0.487341772151899\\
4	0.744725738396624\\
5	0.744725738396624\\
5	0.890295358649789\\
6	0.890295358649789\\
6	0.951476793248945\\
7	0.951476793248945\\
7	0.987341772151899\\
8	0.987341772151899\\
8	1\\
inf	1\\
};
\addlegendentry{G = 100}
\addplot [color=color5, line width=1.5pt, mark=square, mark options={solid}]
  table[row sep=crcr]{%
-inf	0\\
3	0\\
3	0.470897484040556\\
4	0.470897484040556\\
4	0.726248591813744\\
5	0.726248591813744\\
5	0.864814119414195\\
6	0.864814119414195\\
6	0.930529478032294\\
7	0.930529478032294\\
7	0.960195268494179\\
8	0.960195268494179\\
8	0.978971085242208\\
9	0.978971085242208\\
9	0.991363124295907\\
10	0.991363124295907\\
10	0.996620352985355\\
11	0.996620352985355\\
11	0.998873450995118\\
12	0.998873450995118\\
12	1\\
inf	1\\
};
\addlegendentry{G = 1000}
\addplot [color=color4, line width=1.5pt, mark=o, mark options={solid}]
  table[row sep=crcr]{%
0	0\\
1	0.367879441171442\\
2	0.606530659712633\\
3	0.778800783071405\\
4	0.882496902584595\\
5	0.939413062813476\\
6	0.969233234476344\\
7	0.984496437005408\\
8	0.992217938260244\\
9	0.996101369470118\\
10	0.998048781107476\\
11	0.999023914181976\\
12	0.999511837939889\\
};
\addlegendentry{Upper Bound}

\addplot [color=color3, line width=1.5pt, mark=triangle, mark options={solid}]
  table[row sep=crcr]{%
0   0\\
1	1.3887943864964e-11\\
2	3.72665317207867e-06\\
3	0.00193045413622771\\
4	0.0439369336234074\\
5	0.209611387151098\\
6	0.457833361771614\\
7	0.676633846161729\\
8	0.822577562398665\\
9	0.906960617887384\\
10	0.952344799895176\\
11	0.975881550135659\\
12	0.987867172314\\
};
\addlegendentry{Lower Bound}

\end{axis}

\end{tikzpicture}%
}    
    \caption{CDF of the probability of successful collision resolution for different rounds}
    \label{fig:cdf}
\end{figure*}
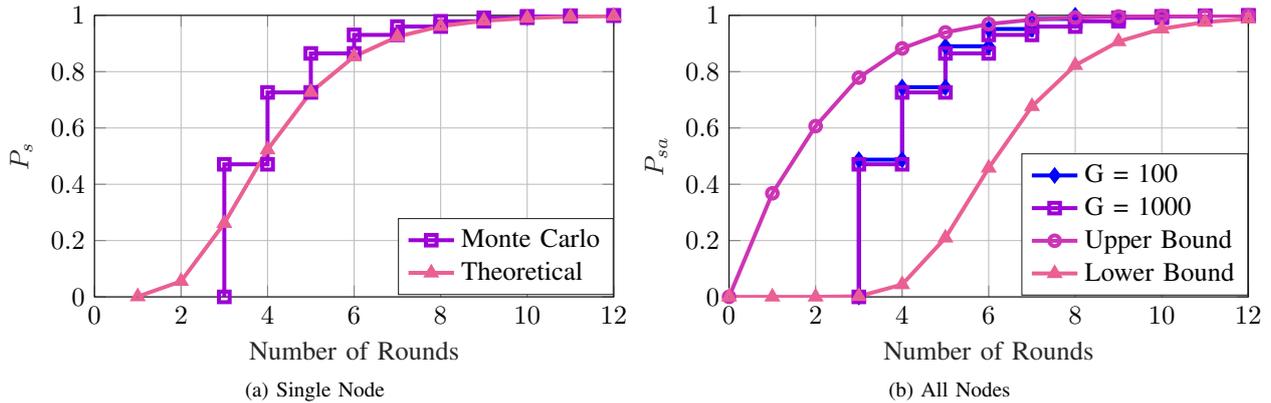
\subsection{Multicast WUR}

In multicast \gls{wur}, the dynamic nature of group selection makes it difficult to perform an analytical study. However, we can give upper and lower bounds to the additional delay and energy consumption introduced by the two collision resolution mechanisms when collisions do happen.

If we consider linear search, the additional energy consumption is simple: since nodes can stop listening after being polled, they spend no more than $GE_{\text{wu}}$ units of energy to listen to the round of wake-up messages, and those that had packets to send spend an additional $E_p$ for each packet. On the other hand, the expected additional delay is at least  $\frac{G(\tau_0+t_{\text{wu}})+2E_p}{2}$, in the best case scenario in which collisions involve only 2 nodes.

In binary search, the worst scenario when a collision has occurred is when all the nodes have the same activation probability, which we denote by $p_a$. In this case, the collision probability can be defined as
\begin{align}
    p_{c}(G) = 1 - (1-p_a)^G - Gp_a(1-p_a)^{G-1}\,.
\end{align}
The probability that $M$ out of the $G$ nodes in the group will be active is then
\begin{align}
    P_a (M|G) = \dfrac{G!}{M!(G-M)!} p_a^M (1-p_a)^{G-M},
\end{align}
and the probability of activation for $M\geq2$ nodes in a group of size $G$ given a collision has occurred is given by
\begin{align}
    P_a (M|G,c) = \dfrac{\frac{G!}{M!(G-M)!} p_a^M (1-p_a)^{G-M}}{1-(1-p_a)^G - Gp_a(1-p_a)^{G-1}},
\end{align}
where $c$ represents the collision event. Using binary search, the group is divided in half at each round until all collisions are resolved, i.e., each subgroup contains no more than one active node. To achieve this, the \gls{gw} has to divide and poll the subgroups in an iterative fashion. Hence, the conditional probability that a given packet is successfully transmitted by the end of the $r$-th round of polling given $M$ active users in a group of $G$ is given by:
\begin{equation}
    P_{s} (r|M,G) = \begin{cases}\prod_{\ell=1}^{M-1} \frac{G_r + 1 - \ell}{G+1-\ell} & \text{if } M \leq G_r + 1; \\
    0 & \text{otherwise,}\\
    \end{cases}
    \label{eq:success_prob}
\end{equation}
where $G_r = G-2^{-r}G$ is the size of the node's polling group at the $r$-th consecutive round of binary search, and the argument of the product is the probability that the $\ell$-th node with a pending packet is placed in a different set than the tagged one. We can use~\eqref{eq:success_prob} and apply the law of total probability to compute the probability of receiving the packet from a particular node at the $r$-th round for a group of size $G$. Additionally, the duration of the $r$-th round depends on the number of remaining collisions, as does the energy consumption. To compute the probability of resolving all collisions by the $r$-th round, we consider that each subgroup should contain only one active node: if two or more sensors are in the same polling group, the collision will recur. The total number of possible combinations of active sensors is given by:
\begin{equation}
C(G,M)=\binom{G}{M}.
\end{equation}
However, we only consider the combinations that result in each sensor being in a different group, whose number is:
\begin{equation}
    S_r(G,M)=\prod_{i=0}^{M-1}G\left(\frac{2^r-i}{2^r}\right)=G^M\prod_{i=0}^{M-1}\left(1-2^{-r}i\right).
\end{equation}
 Considering that $1-2^{-r}i \leq e^{-i/2^r}$, we can approximate and simplify this to get:
\begin{equation}
    P_{sa}(r|M)\leq\frac{G^M(G-M)!M!}{G!}\exp\left(-\frac{M^2}{2^{r+1}}\right),
    \label{eq:collision_success}
\end{equation}
which is a bound on the final probability that all collisions are resolved by the end of the $r$-th round for a given number of active nodes $M$. Note that, in this derivation, we assumed that no other nodes in the group become active, \ie, there are no new packet arrivals during the collision resolution process. Packets arriving during a collision resolution procedure should be buffered and transmitted at the following polling cycle. 

Fig.~\ref{fig:cdf} shows the theoretical bound and Monte Carlo probability of successful collision resolution. \figurename~\ref{fig:cdf_single} shows the probability of successful collision resolution for a single node using~\eqref{eq:success_prob} with group size $G=1000$ and $M=\sum_{n \in \C{G}} \lambda_n \E{\tau_n}$. As visible from the figure, the theoretical analysis slightly underestimates the probability of collision resolution in a particular round compared to Monte Carlo analysis. This is due to the fact that the theoretical analysis conservatively considers a high number of active nodes, while in the Monte Carlo analysis, the number of active nodes is almost always smaller. \figurename~\ref{fig:cdf_all} shows the theoretical number of rounds required by the binary search for group sizes $G={100,1000}$ using optimistic ($M=2$) and pessimistic ($M=\sum_{n \in \C{G}} \lambda_n \E{\tau_n}$) numbers of colliding nodes, computed using~\eqref{eq:collision_success} and compared with the Monte Carlo performance. As the figure shows, the Monte Carlo performance is in between the two theoretical curves. Interesting to notice that more than 80\% of collisions are solved within 6 rounds, both when $G=100$ and $G=1000$. 

\section{System-level Simulation and Results}\label{sec:results}

In order to verify our analysis and compare the proposed multicast \gls{wur} protocol with unicast \gls{wur} and legacy ALOHA, we performed a Monte Carlo simulation. We considered two different scenarios, with $N\in\{100,1000\}$ and the parameters given in Tab. \ref{tb:tab1}. In all scenarios, all the nodes are in the \gls{gw}'s communication range.

\begin{table}[]
\caption{Simulation Parameters}
\centering
\begin{tabular}{lc}
\toprule
Parameter & Value \\
\midrule
Simulator           & MATLAB            \\ 
Number of Nodes 	& $\{100,1000\}$    \\
Polling Time for 1 Node ($t_{\text{wu}}+\tau_0$)  & $15$ ms \\
Transmission Time for 1 packet ($t_p$)   & $1$ ms \\
WUR Energy Consumption & $365\times10^{-9}$ W/s \\
PCR Energy Consumption & $0.1$ W/s \\
Aggregate load ($\xi$) & $\{0.01,0.1,\ldots,0.5\}$ $\text{pkt}/t_p$               \\
Collision Probability Threshold & 5\%                \\
Number of iterations & 1000000               \\
\bottomrule
\end{tabular}
\label{tb:tab1}
\end{table}
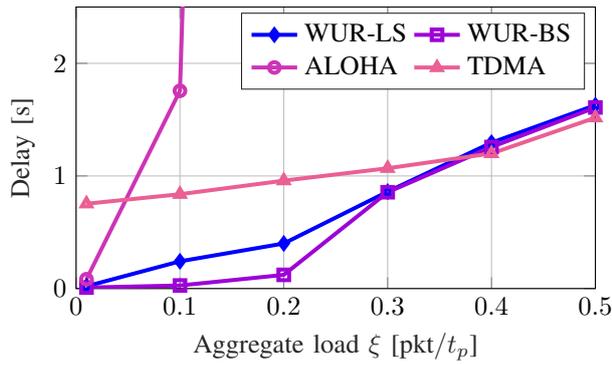
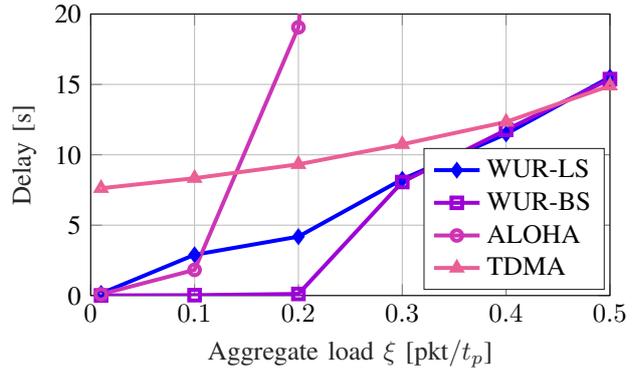
\begin{figure*}[!t]
\centering
\subfloat[%
  $N=100$ nodes. \label{fig:delay_100}%
]{\begin{tikzpicture}
\begin{axis}[%
width=\fwidth,
height=\fheight,
xmin=0,
xmax=0.5,
xlabel style={font=\color{white!15!black}},
xlabel={Aggregate load $\xi$ [pkt$/t_p$]},
ymin=0,
ymax=2.5,
ylabel style={font=\color{white!15!black}},
ylabel={Delay [s]},
xmajorgrids,
ymajorgrids,
legend style={legend cell align=left, legend columns=2, align=left, draw=white!15!black}
]
\addplot [color=color6, line width=1.5pt, mark=diamond, mark options={solid}]
  table[row sep=crcr]{%
0.01	0.0202252294704712\\
0.1	0.241783704435171\\
0.2	0.400125336851459\\
0.3	0.860449611614974\\
0.4	1.2954849518249\\
0.5	1.62839275259259\\
};
\addlegendentry{WUR-LS}

\addplot [color=color5, line width=1.5pt, mark=square, mark options={solid}]
  table[row sep=crcr]{%
0.01	0.00867778575433564\\
0.1	0.0272985624244001\\
0.2	0.121524161268274\\
0.3	0.855242786963021\\
0.4	1.25667553616451\\
0.5	1.60791882483749\\
};
\addlegendentry{WUR-BS}

\addplot [color=color4, line width=1.5pt, mark=o, mark options={solid}]
  table[row sep=crcr]{%
0.01	0.0829058075107737\\
0.1	1.75603956439205\\
0.2	30.7419203606911\\
0.3	107\\
};
\addlegendentry{ALOHA}

\addplot [color=color3, line width=1.5pt, mark=triangle, mark options={solid}]
  table[row sep=crcr]{%
0.01	0.753079527210928\\
0.1	0.83774690012511\\
0.2	0.958247386457432\\
0.3	1.06819486332152\\
0.4	1.19955687064312\\
0.5	1.51741228456523\\
};
\addlegendentry{TDMA}

\end{axis}

\end{tikzpicture}%
}
\subfloat[%
  $N=1000$ nodes. \label{fig:delay_1000}%
]{
%
%
\begin{tikzpicture}

\begin{axis}[%
width=\fwidth,
height=\fheight,
xmin=0,
xmax=0.5,
xlabel style={font=\color{white!15!black}},
xlabel={Aggregate load $\xi$ [pkt$/t_p$]},
ymin=0,
ymax=20,
ylabel style={font=\color{white!15!black}},
ylabel={Delay [s]},
ticks=both,
ytick={0,5,10,15,20},
yticklabels={0,5,10,15,20},
axis background/.style={fill=white},
xmajorgrids,
ymajorgrids,
legend style={legend cell align=left,at={(0.99,0.02)}, anchor=south east, align=left, draw=white!15!black}
]
\addplot [color=color6, line width=1.5pt, mark=diamond, mark options={solid}]
  table[row sep=crcr]{%
0.01	0.148385827867978\\
0.1	2.89744147119019\\
0.2	4.18638976563602\\
0.3	8.25770458512099\\
0.4	11.4824526781798\\
0.5	15.5172963069903\\
};
\addlegendentry{WUR-LS}

\addplot [color=color5, line width=1.5pt, mark=square, mark options={solid}]
  table[row sep=crcr]{%
0.01	0.00911246603779405\\
0.1	0.0276389287708323\\
0.2	0.11384653852593\\
0.3	8.06429781117756\\
0.4	11.7713480233656\\
0.5	15.3787310510476\\
};
\addlegendentry{WUR-BS}

\addplot [color=color4, line width=1.5pt, mark=o, mark options={solid}]
  table[row sep=crcr]{%
0.01	0.0911968348170129\\
0.1	1.82453027981614\\
0.2	19.0549838165002\\
0.3	92\\
};
\addlegendentry{ALOHA}

\addplot [color=color3, line width=1.5pt, mark=triangle, mark options={solid}]
  table[row sep=crcr]{%
0.01	7.62265489500474\\
0.1	8.33563684516496\\
0.2	9.32296377188445\\
0.3	10.743897722434\\
0.4	12.3358890671137\\
0.5	14.90583664012\\
};
\addlegendentry{TDMA}

\end{axis}
\end{tikzpicture}%
}    
    \caption{Average delay of all packets in the network at different data arrival rate}
    \label{fig:delay}
\end{figure*}

The simulation is performed with 1000000 iterations for any given number of nodes and arrival rate. Each node in the network has a random individual arrival rate in each iteration, but we fix the aggregate arrival rate $\xi=\sum_{n=1}^N\lambda_nt_p$: some nodes have a high arrival rate while some have a low arrival rate, but the total average number of packets generated in each transmission duration $t_p$ remains constant.

The energy efficiency $\eta_n$ of node $n$ is given by:
\begin{equation}
    \eta_n = \dfrac{E_{\text{tx}}K_n}{E_n},
\end{equation}
where $K_n$ and $E_n$ represent the total number of packets successfully transmitted by node $n$ and its total energy consumption, respectively, over the whole simulation.

In our evaluation, we consider the following protocols:
\begin{itemize}
    \item \gls{wur}-LS: \gls{wur} multicast polling with collision resolution using linear search;
    \item \gls{wur}-BS: \gls{wur} multicast polling with collision resolution using binary search;
    \item ALOHA: legacy slotted ALOHA protocol; as soon as nodes have data, they transmit at the next possible slot, with retransmission after a random backoff in case of collisions;
    \item TDMA: Unicast \gls{wur} polling with round robin scheduling.
\end{itemize}

\begin{figure*}[!t]
\centering
\subfloat[%
  $N=100$ nodes. \label{fig:energy_100}%
]{
%
%
\begin{tikzpicture}

\begin{axis}[%
width=\fwidth,
height=\fheight,
xmin=0,
xmax=0.5,
xlabel style={font=\color{white!15!black}},
xlabel={Aggregate load [pkt/$t_p$]},
ymin=40,
ymax=100,
ylabel style={font=\color{white!15!black}},
ylabel={Energy efficiency $\eta$ [\%]},
axis background/.style={fill=white},
xmajorgrids,
ymajorgrids,
legend style={legend cell align=left, legend columns=2,align=left, draw=white!15!black}
]
\addplot [color=color6, line width=1.5pt, mark=diamond, mark options={solid}]
  table[row sep=crcr]{%
0.01	97.2058441939026\\
0.1	64.0915889240527\\
0.2	52.6337103424245\\
0.3	54.4826384870733\\
0.4	62.0300030535238\\
0.5	78.8258722757895\\
};
\addlegendentry{WUR-LS}

\addplot [color=color5, line width=1.5pt, mark=square, mark options={solid}]
  table[row sep=crcr]{%
0.01	97.4887908613785\\
0.1	72.005765120335\\
0.2	47.816675361304\\
0.3	66.1243673712746\\
0.4	77.1950920247372\\
0.5	83.4722494998828\\
};
\addlegendentry{WUR-BS}

\addplot [color=color4, line width=1.5pt, mark=o, mark options={solid}]
  table[row sep=crcr]{%
0.01	99.1051454138702\\
0.1	88.7684694112847\\
0.2	72.7824483454575\\
0.3	56.7655127542978\\
0.4	56.7577548005908\\
0.5	56.8112942567817\\
};
\addlegendentry{ALOHA}

\addplot [color=color3, line width=1.5pt, mark=triangle, mark options={solid}]
  table[row sep=crcr]{%
0.01	100\\
0.1	100\\
0.2	100\\
0.3	100\\
0.4	100\\
0.5	100\\
};
\addlegendentry{TDMA}

\end{axis}
\end{tikzpicture}%
}
\subfloat[%
  $N=1000$ nodes. \label{fig:energy_1000}%
]{
%
%
\begin{tikzpicture}

\begin{axis}[%
width=\fwidth,
height=\fheight,
xmin=0,
xmax=0.5,
xlabel style={font=\color{white!15!black}},
xlabel={Aggregate load [pkt$/t_p$]},
ymin=40,
ymax=100,
ylabel style={font=\color{white!15!black}},
ylabel={Energy efficiency $\eta$ [\%]},
axis background/.style={fill=white},
xmajorgrids,
ymajorgrids,
legend style={legend cell align=left, legend columns=2,align=left, draw=white!15!black}
]
\addplot [color=color6, line width=1.5pt, mark=diamond, mark options={solid}]
  table[row sep=crcr]{%
0.01	97.3402486400332\\
0.1	53.527870476145\\
0.2	59.4325927814383\\
0.3	61.8212694126564\\
0.4	59.9464143153429\\
0.5	47.9446752032704\\
};
\addlegendentry{WUR-LS}

\addplot [color=color5, line width=1.5pt, mark=square, mark options={solid}]
  table[row sep=crcr]{%
0.01	97.1755897681225\\
0.1	71.8265687733092\\
0.2	47.8042516223864\\
0.3	64.7407835611172\\
0.4	74.4550071110744\\
0.5	80.7164158770994\\
};
\addlegendentry{WUR-BS}

\addplot [color=color4, line width=1.5pt, mark=o, mark options={solid}]
  table[row sep=crcr]{%
0.01	99.0205680705191\\
0.1	88.4112216162264\\
0.2	74.2811418363169\\
0.3	58.1289101900624\\
0.4	57.2804479451321\\
0.5	57.2763747618153\\
};
\addlegendentry{ALOHA}

\addplot [color=color3, line width=1.5pt, mark=asterisk, mark options={solid}]
  table[row sep=crcr]{%
0.01	100\\
0.1	100\\
0.2	100\\
0.3	100\\
0.4	100\\
0.5	100\\
};
\addlegendentry{TDMA}

\end{axis}

\end{tikzpicture}%
}
    \caption{Average Energy efficiency achieved in the network at different data arrival rate}
    \label{fig:energy}
\end{figure*}
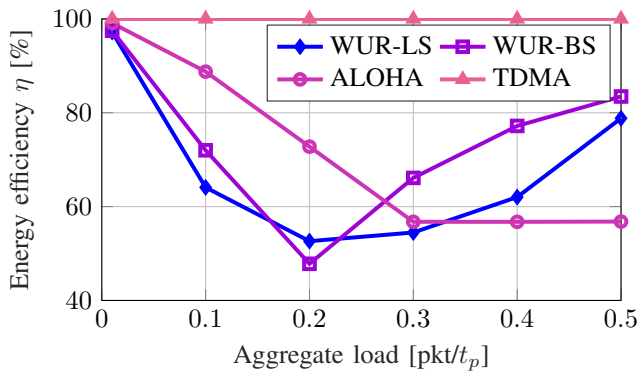
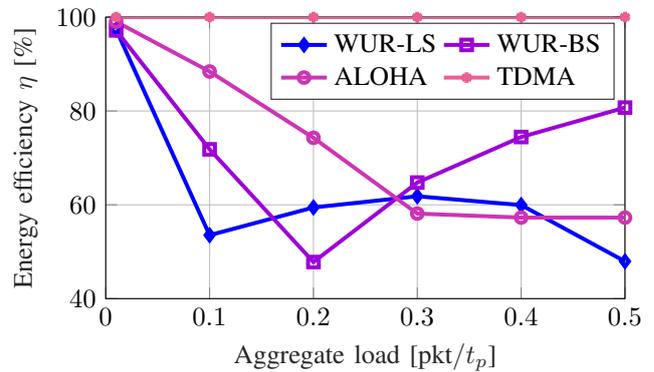

The two metrics that we consider in our performance evaluation are the average delay and energy efficiency. These metrics are most important in \gls{iot} applications such as tracking and monitoring in remote places. 

\figurename~\ref{fig:delay} shows the delay performance of the selected schemes as a function of the total traffic load, considering $N=100$ and $N=1000$ nodes in the network, respectively.
As the figure shows, the delay incurred by the \gls{wur}-LS and \gls{wur}-BS is lower than that attained using ALOHA in all cases, and the latter is unstable even for relatively low aggregate load levels. On the other hand, \gls{tdma} achieves a slightly lower delay if the load is extremely high (at least 0.5 pkt$/t_p$), as explicit coordination becomes the only way to manage all the generated traffic. If the load is low, the waiting time between subsequent wake-up signals for a single node is the dominant factor contributing to delay in \gls{tdma}, while collisions in ALOHA become unsustainable even for relatively low load levels. We can also note that binary search is significantly more effective at reducing the delay with respect to linear search, but only for low load levels ($\leq 0.3$ pkt$/t_p$): as the number of collided packets increases, binary search becomes less effective, requiring more and more rounds to solve all collisions. 

However, multicast polling has a cost: both schemes have a lower energy efficiency with respect to \gls{tdma} and even ALOHA, consuming almost double the energy of \gls{tdma} with unicast polling, as \figurename~\ref{fig:energy} shows. The large number of polls required by linear search collision resolution makes it generally worse than binary search in terms of energy efficiency: as the number of collisions increases, linear search tends to suffer from the same snowball effect as ALOHA (the time necessary to resolve a collision leads to a backlog of packets, which further increases the probability of future collisions), while binary search can avoid it for longer. As the aggregate load increases, both multicast schemes also gradually reduce the size of the polling groups to maintain an acceptable collision probability. For this reason, higher loads always correspond to a higher delay, but energy efficiency actually improves for binary search if the load is significant. 

\begin{figure*}[!t]
\centering
\subfloat[%
  WUR-BS delay gain. \label{fig:extra_delay}%
]{\begin{tikzpicture}

\begin{axis}[%
width=\fwidth,
height=\fheight,
xmin=0,
xmax=1000,
xlabel style={font=\color{white!15!black}},
xlabel={Group Size},
ymin=-10,
ymax=20,
ylabel style={font=\color{white!15!black}},
ylabel={$\Delta D\text{ [s]}$},
axis background/.style={fill=white},
xmajorgrids,
ymajorgrids,
legend style={at={(0.01,0.98)}, legend columns=2,anchor=north west}
]
\addplot [color=color1, line width=1.5pt, mark=triangle, mark options={solid}]
  table[row sep=crcr]{%
1	-0.0899815709485772\\
100	-0.0422507550218576\\
200	-0.0100990339707998\\
300	0.0901125204423368\\
400	0.0376090348332601\\
500	-0.00655767886170988\\
600	0.0594605977819112\\
700	0.0505414177606489\\
800	0.0832905203587352\\
900	0.105701912354839\\
1000	0.15361786307011\\
};
\addlegendentry{$\xi=0.01$}

\addplot [color=color2, line width=1.5pt, mark=o, mark options={solid}]
  table[row sep=crcr]{%
1	-0.0291991259635651\\
100	0.0106503222839329\\
200	0.0235616149320679\\
300	-0.0188349590818495\\
400	0.0909432119041469\\
500	0.202913395596866\\
600	0.285827477460946\\
700	0.594307109625843\\
800	0.918279599533655\\
900	1.32667833870352\\
1000	1.78363057411999\\
};
\addlegendentry{$\xi=0.1$}

\addplot [color=color3, line width=1.5pt, mark=square, mark options={solid}]
  table[row sep=crcr]{%
1	0.0224393739716717\\
100	-0.0705543716631389\\
200	-0.0373538311743431\\
300	0.066465050792524\\
400	0.122008066975294\\
500	0.233360375074835\\
600	0.700002815729814\\
700	1.25271925772208\\
800	1.9833459436594\\
900	3.14407801411808\\
1000	4.27774963585861\\
};
\addlegendentry{$\xi=0.2$}

\addplot [color=color4, line width=1.5pt, mark=diamond, mark options={solid}]
  table[row sep=crcr]{%
1	-0.189045002157378\\
100	-0.0282858685872043\\
200	-0.137274502380272\\
300	-0.240931390253206\\
400	-0.118367110978996\\
500	0.133422091158278\\
600	0.654829473590558\\
700	1.64570132256318\\
800	3.29621351164107\\
900	5.09722738653208\\
1000	7.4936639345594\\
};
\addlegendentry{$\xi=0.3$}

\addplot [color=color5, line width=1.5pt, mark=Mercedes star, mark options={solid}]
  table[row sep=crcr]{%
1	-0.00803807094389519\\
100	-0.0923462520869975\\
200	-0.488204450852422\\
300	-0.848876324648863\\
400	-1.24901274773391\\
500	-1.32290795191399\\
600	-0.162843855271509\\
700	1.20558217021987\\
800	4.01956899422502\\
900	7.71209182877121\\
1000	11.21847913195\\
};
\addlegendentry{$\xi=0.4$}

\addplot [color=color6, line width=1.5pt, mark=x, mark options={solid}]
  table[row sep=crcr]{%
1	0.239555931876225\\
100	-0.359668505915797\\
200	-1.4729505089838\\
300	-2.82361398236291\\
400	-4.11295305071074\\
500	-4.20942910100691\\
600	-6.0417967555328\\
700	-3.61650226008851\\
800	1.59843741246875\\
900	9.35162494434747\\
1000	16.3364287916423\\
};
\addlegendentry{$\xi=0.5$}

\end{axis}

\end{tikzpicture}%
}
\subfloat[%
  WUR-BS energy efficiency loss. \label{fig:extra_energy}%
]{
%
%
\begin{tikzpicture}

\begin{axis}[%
width=\fwidth,
height=\fheight,
xmin=0,
xmax=1000,
xlabel style={font=\color{white!15!black}},
xlabel={Group Size},
ymin=0,
ymax=30,
ylabel style={font=\color{white!15!black}},
ylabel={$\Delta \eta~[\%]$},
axis background/.style={fill=white},
xmajorgrids,
ymajorgrids,
legend style={at={(0.01,0.98)}, legend columns=3,anchor=north west}
]
\addplot [color=color1, line width=1.5pt, mark=triangle, mark options={solid}]
  table[row sep=crcr]{%
1	0.00272975618372806\\
100	0.0557240080845012\\
200	0.273891826777617\\
300	0.406101907839362\\
400	0.80414960865689\\
500	-0.0225519800414298\\
600	0.483755589334223\\
700	0.738293722196881\\
800	2.45917739741957\\
900	2.04542486509188\\
1000	2.13551054283044\\
};
\addlegendentry{$\xi=0.01$}

\addplot [color=color2, line width=1.5pt, mark=o, mark options={solid}]
  table[row sep=crcr]{%
1	0.00631038928531336\\
100	1.49142360425977\\
200	1.54566397103302\\
300	3.1448676716766\\
400	4.59386989720689\\
500	7.59233443479563\\
600	9.64294114413278\\
700	8.21732342621404\\
800	11.4946916553798\\
900	13.722923844045\\
1000	14.9477014464947\\
};
\addlegendentry{$\xi=0.1$}

\addplot [color=color3, line width=1.5pt, mark=square, mark options={solid, rotate=90}]
  table[row sep=crcr]{%
1	0.00156133143077719\\
100	1.50815011794838\\
200	3.21365966731927\\
300	5.19666793964635\\
400	7.09766532879209\\
500	11.640709377582\\
600	12.1106663427359\\
700	15.9927930647425\\
800	18.3700628627014\\
900	19.0387357468905\\
1000	21.347938751601\\
};
\addlegendentry{$\xi=0.2$}

\addplot [color=color4, line width=1.5pt, mark=diamond, mark options={solid}]
  table[row sep=crcr]{%
1	0.00828386674224246\\
100	1.7369413075826\\
200	2.85716815281856\\
300	5.40807166760107\\
400	9.14743571990789\\
500	12.1443337785812\\
600	14.8060843165665\\
700	17.2643526201308\\
800	18.9716020359181\\
900	20.399788561722\\
1000	21.9606454622898\\
};
\addlegendentry{$\xi=0.3$}

\addplot [color=color5, line width=1.5pt, mark=Mercedes star, mark options={solid}]
  table[row sep=crcr]{%
1	-0.0120110573234578\\
100	2.11562145439494\\
200	4.73639075080607\\
300	8.32292269799629\\
400	11.8239195793152\\
500	15.8782474163781\\
600	16.1627047563538\\
700	18.4571280315827\\
800	19.5979495605257\\
900	19.8558726847165\\
1000	20.6664144953219\\
};
\addlegendentry{$\xi=0.4$}

\addplot [color=color6, line width=1.5pt, mark=x, mark options={solid}]
  table[row sep=crcr]{%
1	0.0180901444510995\\
100	1.63883679624252\\
200	6.00610514193044\\
300	10.2161690382591\\
400	15.5286694472482\\
500	16.7323957509906\\
600	19.3127197262275\\
700	18.433065745335\\
800	17.8250772071579\\
900	17.7548918403172\\
1000	17.1729905585398\\
};
\addlegendentry{$\xi=0.5$}

\end{axis}

\end{tikzpicture}%
}
    \caption{Performance difference between WUR-LS and WUR-BS with a fixed group size and 1000 total nodes.}
    \label{fig:groups}
\end{figure*}
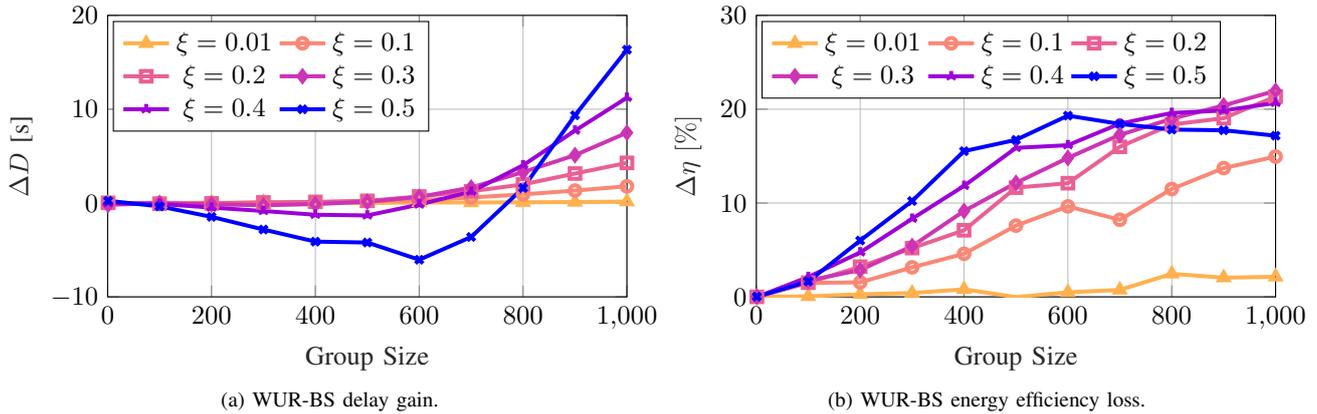

Additionally, to understand the difference in performance obtained by using binary search or linear search for fast collision resolution, we evaluate the \gls{wur}-LS and \gls{wur}-BS schemes for fixed group size: instead of dynamic-sized groups that keep the collision probability below a certain threshold, we consider multicast polls of a fixed group consisting of the nodes which have the highest probability of having a packet. \figuresname~\ref{fig:extra_delay} and \ref{fig:extra_energy} show the extra delay and difference in energy efficiency encountered when considering the \gls{wur}-BS scheme compared to \gls{wur}-LS scheme. The extra delay and energy efficiency difference is given by
\begin{align}
    \Delta D &= D^{\text{WUR-LS}} - D^{\text{WUR-BS}};\\
    \Delta \eta &= \eta^{\text{WUR-LS}} - \eta^{\text{WUR-BS}} . 
\end{align}

Fig.~\ref{fig:extra_delay} shows that linear search is more effective for medium-sized groups when the traffic load is high, but larger groups tend to favor binary search even when the aggregate load is high. If groups are very small, the two schemes are approximately similar in their effectiveness, but the logarithmically increasing delay of binary search is significantly lower than the linearly increasing delay of linear search when the group size grows.

At the same time, the energy efficiency of binary search is much lower than for linear search, as the possibility of collisions in later rounds requires a significantly higher expense than just a series of polls. This difference is significant, as subsequent collisions are much more expensive than individual polls. However, the energy efficiency of the multicast schemes also depends on the group sizes and delay: as each collision is solved in a much shorter time (unless the traffic is very high), binary search allows the receiver to poll fewer nodes in each group with better delay performance, avoiding the energy issue, as the previously discussed results showed.

\section{Conclusions and Future Work}\label{sec:conc}

In this work, we propose and investigate an alternative polling scheme for \gls{wur}. Instead of polling each individual node, our scheme uses multicast polls to wake up multiple nodes at once, striking a balance between the long waiting times of \gls{tdma}-like approaches and the frequent collisions of ALOHA. The resulting scheme is slightly more energy-intensive than either, but can significantly reduce delay in scenarios with relatively light traffic.

Possible avenues of future work may include considering different access priorities for nodes and/or data, multihop transmissions, or \glsfirst{aoi} approaches. Additionally, considering out of band \gls{wur} can decouple data transmission from \gls{wur} signals, possibly reducing polling delays. Another interesting avenue to pursue can be the use of machine learning to predict (potentially correlated) arrival rates based on previous arrivals, which can be then be used to devise multicast groups proactively. 
\section*{Acknowledgments}

This work was supported by the European Union's NextGenerationEU instrument, under the Italian National Recovery and Resilience Plan (NRRP) of NextGenerationEU, as part of the partnership on ``Telecommunications of the Future'' (PE0000001 - program
``RESTART''), the CHIST-ERA-19-CES-002 project "ANDROMEDA",  and the Young Researchers grant SoE0000009 ``REDIAL.''

\bibliographystyle{IEEEtran}
\bibliography{WURbibiography}

\begin{thebibliography}{10}
\providecommand{\url}[1]{#1}
\csname url@samestyle\endcsname
\providecommand{\newblock}{\relax}
\providecommand{\bibinfo}[2]{#2}
\providecommand{\BIBentrySTDinterwordspacing}{\spaceskip=0pt\relax}
\providecommand{\BIBentryALTinterwordstretchfactor}{4}
\providecommand{\BIBentryALTinterwordspacing}{\spaceskip=\fontdimen2\font plus
\BIBentryALTinterwordstretchfactor\fontdimen3\font minus
  \fontdimen4\font\relax}
\providecommand{\BIBforeignlanguage}[2]{{%
\expandafter\ifx\csname l@#1\endcsname\relax
\typeout{** WARNING: IEEEtran.bst: No hyphenation pattern has been}%
\typeout{** loaded for the language `#1'. Using the pattern for}%
\typeout{** the default language instead.}%
\else
\language=\csname l@#1\endcsname
\fi
#2}}
\providecommand{\BIBdecl}{\relax}
\BIBdecl

\bibitem{wang2021evolution}
J.~Wang, M.~K. Lim, C.~Wang, and M.-L. Tseng, ``The evolution of the {Internet
  of Things (IoT)} over the past 20 years,'' \emph{Computers \& Industrial
  Engineering}, vol. 155, p. 107174, 2021.

\bibitem{zanella2023iot}
A.~Zanella, S.~Zubelzu, M.~Bennis, M.~Capuzzo, and P.~Tarolli, ``Internet of
  things for hydrology: Potential and challenges,'' in \emph{Wireless On-demand
  Network systems and Services Conference (WONS 2023)}.\hskip 1em plus 0.5em
  minus 0.4em\relax IEEE, 2023.

\bibitem{georgiou2017iot}
K.~Georgiou, S.~Xavier-de Souza, and K.~Eder, ``The {IoT} energy challenge: A
  software perspective,'' \emph{IEEE Embedded Systems Letters}, vol.~10, no.~3,
  pp. 53--56, 2017.

\bibitem{8030484}
D.~Zucchetto and A.~Zanella, ``Uncoordinated access schemes for the {IoT}:
  Approaches, regulations, and performance,'' \emph{IEEE Communications
  Magazine}, vol.~55, no.~9, pp. 48--54, 2017.

\bibitem{yu2020stabilizing}
J.~Yu, P.~Zhang, L.~Chen, J.~Liu, R.~Zhang, K.~Wang, and J.~An, ``Stabilizing
  frame slotted {ALOHA}-based {IoT} systems: A geometric ergodicity
  perspective,'' \emph{IEEE Journal on Selected Areas in Communications},
  vol.~39, no.~3, pp. 714--725, 2020.

\bibitem{levy1990polling}
H.~Levy and M.~Sidi, ``Polling systems: applications, modeling, and
  optimization,'' \emph{IEEE Transactions on Communications}, vol.~38, no.~10,
  pp. 1750--1760, 1990.

\bibitem{deng2020ieee}
D.-J. Deng, S.-Y. Lien, C.-C. Lin, M.~Gan, and H.-C. Chen, ``{IEEE} 802.11 ba
  wake-up radio: Performance evaluation and practical designs,'' \emph{IEEE
  Access}, vol.~8, pp. 141\,547--141\,557, 2020.

\bibitem{tang2017energy}
S.~Tang and S.~Obana, ``Energy efficient downlink transmission in wireless
  {LANs} by using low-power wake-up radio,'' \emph{Wireless Communications and
  Mobile Computing}, no. 2405381, 2017.

\bibitem{rostami2019wake}
S.~Rostami, S.~Lagen, M.~Costa, M.~Valkama, and P.~Dini, ``Wake-up radio based
  access in {5G} under delay constraints: Modeling and optimization,''
  \emph{IEEE Transactions on Communications}, vol.~68, no.~2, pp. 1044--1057,
  2019.

\bibitem{chiariotti2022scheduling}
F.~Chiariotti, A.~E. Kal{\o}r, J.~Holm, B.~Soret, and P.~Popovski, ``Scheduling
  of sensor transmissions based on {Value of Information} for summary
  statistics,'' \emph{IEEE Networking Letters}, vol.~4, no.~2, pp. 92--96,
  2022.

\bibitem{shiraishi2022query}
J.~Shiraishi, F.~Chiariotti, I.~Leyva-Mayorga, P.~Popovski, H.~Yomo
  \emph{et~al.}, ``Query timing analysis for content-based wake-up realizing
  informative {IoT} data collection,'' \emph{IEEE Wireless Communications
  Letters}, 2022.

\bibitem{feeney2001investigating}
L.~M. Feeney and M.~Nilsson, ``Investigating the energy consumption of a
  wireless network interface in an ad hoc networking environment,'' in
  \emph{Conference on Computer Communications (INFOCOM)}, vol.~3.\hskip 1em
  plus 0.5em minus 0.4em\relax IEEE, 2001, pp. 1548--1557.

\bibitem{mahlknecht2009wur}
S.~Mahlknecht and M.~S. Durante, ``{WUR-MA}c: energy efficient wakeup receiver
  based {MAC} protocol,'' \emph{IFAC Proceedings Volumes}, vol.~42, no.~3, pp.
  79--83, 2009.

\bibitem{bankov2019ieee}
D.~Bankov, E.~Khorov, A.~Lyakhov, and E.~Stepanova, ``{IEEE} 802.11
  ba—extremely low power {Wi-Fi} for massive {Internet of
  Things}—challenges, open issues, performance evaluation,'' in
  \emph{International Black Sea Conference on Communications and Networking
  (BlackSeaCom)}.\hskip 1em plus 0.5em minus 0.4em\relax IEEE, 2019.

\bibitem{mazloum2011dcw}
N.~S. Mazloum and O.~Edfors, ``{DCW-MAC}: An energy efficient medium access
  scheme using duty-cycled low-power wake-up receivers,'' in \emph{Vehicular
  Technology Conference (VTC Fall)}.\hskip 1em plus 0.5em minus 0.4em\relax
  IEEE, 2011.

\bibitem{le2015asynchronous}
T.~N. Le, A.~Pegatoquet, and M.~Magno, ``Asynchronous on demand {MAC} protocol
  using wake-up radio in wireless body area network,'' in \emph{6th
  International Workshop on Advances in Sensors and Interfaces (IWASI)}.\hskip
  1em plus 0.5em minus 0.4em\relax IEEE, 2015, pp. 228--233.

\bibitem{ullah2012energy}
N.~Ullah, M.~S. Chowdhury, M.~Al~Ameen, and K.~S. Kwak, ``Energy efficient
  {MAC} protocol for low-energy critical infrastructure monitoring networks
  using wakeup radio,'' \emph{International Journal of Distributed Sensor
  Networks}, vol.~8, no.~4, p. 504946, 2012.

\bibitem{ait2016opwum}
F.~Ait~Aoudia, M.~Gautier, and O.~Berder, ``{OPWUM}: Opportunistic {MAC}
  protocol leveraging wake-up receivers in {WSNs},'' \emph{Journal of sensors},
  vol. 2016, 2016.

\bibitem{oller2015has}
J.~Oller, I.~Demirkol, J.~Casademont, J.~Paradells, G.~U. Gamm, and L.~Reindl,
  ``Has time come to switch from duty-cycled {MAC} protocols to wake-up radio
  for wireless sensor networks?'' \emph{IEEE/ACM Transactions on Networking},
  vol.~24, no.~2, pp. 674--687, 2015.

\bibitem{hwang2018wake}
S.~Hwang, I.~Kim, K.-M. Kang, and S.~Park, ``Wake-up latency evaluation of
  {IEEE 802.11 ba WUR} system,'' in \emph{International Conference on
  Information and Communication Technology Convergence (ICTC)}.\hskip 1em plus
  0.5em minus 0.4em\relax IEEE, 2018, pp. 880--882.

\end{thebibliography}

\end{document}